\documentclass[journal,draftclsnofoot,onecolumn,12pt,twoside]{IEEEtran}
\usepackage{amsmath,amsfonts}
\usepackage{algorithmic}
\usepackage{algorithm}
\usepackage{array}
\usepackage[caption=false,font=normalsize,labelfont=sf,textfont=sf]{subfig}
\usepackage{textcomp}
\usepackage{stfloats}
\usepackage{url}
\usepackage{verbatim}
\usepackage{graphicx}
\usepackage{cite}
\usepackage{balance}
\usepackage{multirow}
\usepackage{color}
\begin{document}

\title{{Free Space Optical Communication for Inter- Satellite Link: Architecture, Potentials and Trends}

\author{Guanhua Wang,
Fang Yang,~\IEEEmembership{Senior Member,~IEEE},
Jian Song,~\IEEEmembership{Fellow,~IEEE},
and Zhu Han,~\IEEEmembership{Fellow,~IEEE}}
\thanks{This work was supported by the National Key Research and Development Program of China under Grant 2022YFE0101700. (\textit{Corresponding author: Fang Yang.})}
\thanks{Guanhua Wang, Fang Yang and Jian Song are with the Department of Electronic Engineering, Beijing National Research Center for Information Science and Technology, Tsinghua University, Beijing 10084, P. R. China(email: wanggh21@mails.tsinghua.edu.cn, fangyang@tsinghua.edu.cn, jsong@tsinghua.edu.cn).}
\thanks{Fang Yang and Jian Song are also with Key Laboratory of Digital TV System of Guangdong Province and Shenzhen City, Shenzhen 518057, P. R. China.}
\thanks{Zhu Han is with the Electrical and Computer Engineering Department, University of Houston, Houston, TX 77004 USA, and also with the Department of Computer Science and Engineering, Kyung Hee University, Seoul 446-701, South Korea(email: zhan2@uh.edu)}}
\maketitle

\begin{abstract}
The sixth-generation (6G) network is expected to achieve global coverage based on the space-air-ground integrated network, and the latest satellite network will play an important role in it. The introduction of inter-satellite links (ISLs) can significantly improve the throughput of the satellite network, and recently gets lots of attention from both academia and industry. In this paper, we illustrate the advantages of using the laser for ISLs due to its longer communication distance, higher data speed, and stronger security. Specifically, space-borne laser terminals with the acquisition, pointing and tracking mechanism which realize long-distance communication are illustrated, advanced modulation and multiplexing modes that make high communication rates possible are introduced, and the security of ISLs ensured by the characteristics of both laser and the optical channel is also analyzed. Moreover, some open issues such as advanced optical beam steering, routing and scheduling algorithm, and integrated sensing and communication are discussed to direct future research.
\end{abstract}

\section{Introduction}
\IEEEPARstart{T}{he} terrestrial base stations (BSs) play an important role in the current wireless network, however, limited by the economic benefits and transmission distance, BSs have difficulty in achieving coverage of remote areas, ocean, and sky. Nevertheless, satellite communication can achieve global coverage because of the high altitude. Traditional satellite communication systems such as Iridium and Globalstar have been developed since the 1990s. The throughput of traditional satellite systems is difficult to compare with terrestrial communication because of their high cost, limited number, and scarce spectrum. With the development of space technology, satellite communication systems in recent years such as Starlink, Telesat, and OneWeb utilize amounts of low Earth orbit (LEO) satellites with restricted size, weight, and power (SWaP). Small satellites are easy to be substituted and multiple satellites can be launched within one rocket, which makes a large-scale constellation with thousands of satellites possible. Therefore, a constellation with a large number of satellites makes the system more robust and has higher throughput upto Tbps. The number of satellites, constellation architecture, and throughput of Telesat, OneWeb, and Starlink were compared in~\cite{LEOCompare}. The throughput of Telesat with inter-satellite links (ISLs) can be close to OneWeb with less than half the satellites, because the ISLs allow information to be transferred between satellites directly instead of being relayed by ground stations~\cite{LEOCompare}, which implies that the ISLs can improve the performance of the system significantly. 

\begin{figure*}
    \centering
    \includegraphics[width=7in]{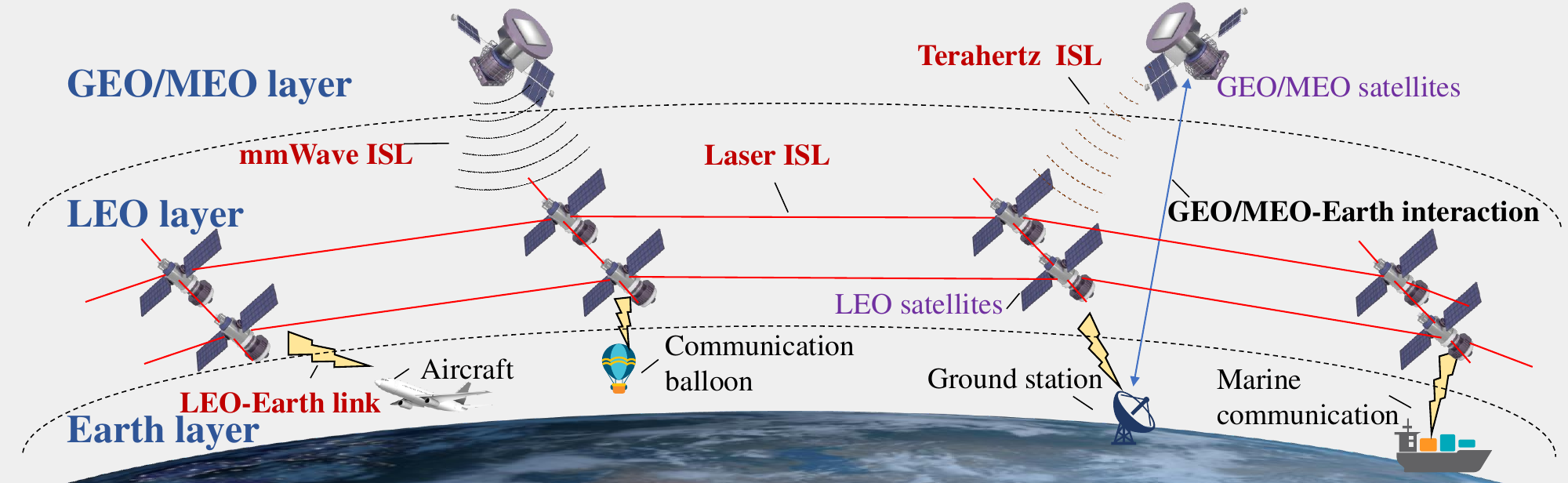}
    \caption{The architecture of satellite network with different types of ISLs.}
    \label{netstructe}
\end{figure*}

A variety of bands can be selected for ISLs, as shown in Fig.~\ref{netstructe}. The millimeter-wave (mmWave) has been utilized in 5G, and the devices and processing algorithms are well researched. However, its bandwidth is insufficient for future satellite communication. Meanwhile, the beam width is positively correlated with the wavelength, and so the mmWave beam is wide. The Terahertz has huge bandwidth, which makes communication upto Tbps possible, and in addition the beam width is appropriate for inter-satellite communication. However, the energy efficiency of current Terahertz devices still needs to be improved, and it is not suitable for satellites with limited power budgets. The spectrum resources in the optical band are rich and unlicensed, and the intensity modulation direct detection (IM-DD) based on photons counting can effectively resist Doppler shift, while the coherent modulation requires countermeasures against Doppler shift through digital signal processing. With high efficiency and long service life, light-emitting diodes (LEDs) have wide applications in visible light communication (VLC). Nonetheless, LEDs emit light beams that are diffuse and incoherent, and have long response time, making it difficult to make full use of the spectrum. On the contrary, lasers have good monochromaticity, high energy concentration, and strong directivity, while the response time is also short, allowing wide-band modulation. The solid-state lasers have high output power and high beam quality, but the disadvantages are low energy efficiency and high heat generation. Laser diodes (LDs) are power efficient and easy to be integrated on a chip, but the output beam quality is not sufficiently good and needs to be collimated through an optical system. In addition, a new type of LD called vertical-cavity surface-emitting laser (VCSEL) can improve the beam quality. These transmission approaches are compared in Table \uppercase\expandafter{\romannumeral1}.

At present, the laser ISLs has attracted attention and started to be applied. In the Starlink constellation, each satellite has four ISLs, two of which are connected to the neighbor satellites in the same orbital plane (OP), and others are connected to the satellites in the neighbor OPs~\cite{interlink}. The velocity of LEO satellites can reach 7 km/s, which puts forward high requirements on accurate pointing and swift tracking for space-borne transceiver terminals. Meanwhile, the range of laser ISLs is upto thousands of kilometers, and thus laser modulation and multiplexing schemes are vital. Furthermore, ensuring the security of ISLs by the characteristics of the laser and the space channels is also worth discussing.

\begin{table*}[t]
    \centerline { TABLE I. Comparison among different signal sources}
    \vskip5pt
    \centering
    \tiny
    \scalebox{1}{
\renewcommand{\arraystretch}{1.5}
\begin{tabular}{c|c|c|c|c|c|c|c|c}
\hline
\hline
\textbf{Emission source type} & \textbf{Spectrum}               & \textbf{Wavelength}           & \textbf{Bandwidth} & \textbf{Beam Width} & \textbf{Power Efficiency} & \textbf{Device Size} & \textbf{Modulation Rate} & \textbf{Doppler Shift} \\ \hline
\textbf{mmWave}               & 30 GHz-300 GHz                    & 1 mm-10 mm                      & $\sim$GHz          & wide                & high                     & medium                 & \textless Gbps           & not resistant          \\ \hline
\textbf{Terahertz}            & 300 GHz-30 THz                    & 10$\mu$m-1 mm                      & $\sim$THz          & narrow              & low                      & medium               & $\sim$Gbps               & not resistant          \\ \hline
\textbf{LED}                  & \multirow{3}{*}{200 THz-1200 THz} & \multirow{3}{*}{250 nm-1550 nm} & $\sim$100 THz       & wide                & high                     & small                & \textless Gbps           & resistant              \\ \cline{1-1} \cline{4-9} 
\textbf{Solid-state laser}    &                                 &                               & $\sim$100 THz       & very narrow         & medium                   & small                & $\sim$10 Gbps             & resistant              \\ \cline{1-1} \cline{4-9} 
\textbf{LD}                   &                                 &                               & $\sim$100 THz       & ordinary            & high                     & small                & $\sim$10 Gbps             & resistant              \\ \hline
\end{tabular}}
\end{table*}

The contributions of this paper are summarized as follows. First, a typical architecture for satellites with ISLs is summarized, and the roles and interrelations of each level are analyzed. Then, the merits of laser ISLs are analyzed in three aspects compared with other bands: extending the communication distance, enhancing the communication rate, and ensuring the communication security. Finally, the future trends and some open issues are discussed, with respect to the latest device technology, machine learning in the scheduling algorithm, and multi-function integration, while some preliminary ideas on these issues are presented.

\section{Satellite Network System Architecture}
The structure of the satellite network is shown in Fig.~\ref{netstructe}. The model is divided into three layers from top to bottom according to altitude: the geostationary orbit (GEO)/medium Earth Orbit (MEO) layer, the LEO layer where the satellites communicate with each other by ISLs, and the Earth layer.

The satellites in the LEO layer are connected by ISLs in a mesh and realize the network access by receiving and transmitting the signal sent from the Earth layer. The LEO satellites serve as a relay for transmitting communication information, whose propagation paths are shorter than satellites in higher OPs because of their low altitudes. Moreover, electromagnetic waves travel faster in space than that in the fiber, which indicates that although the signal transmitted over ISLs travels longer distances, the delay will be less than that of fiber optic communication when the propagation distance is sufficiently long, and it is significant for finance, telemedicine, and other delay-sensitive business. Due to the communication rate demand and the limited SWaP of LEO satellites, laser ISLs are expected to be applied in the LEO layer.

The GEO/MEO layer realizes the collection of state information of the LEO satellites, generates instructions, and sends them back to the LEO layer, while messages with high delay tolerance can also be delivered by this layer as well. High altitude allows for wider coverage of satellites. For instance, only three GEO satellites are needed to cover all the LEO satellites in Starlink. A GEO/MEO satellite can send commands to a large number of LEO satellites at the same time, and the delay requirement of the controlling information is not high because the link status of the whole constellation is updated every dozens of seconds. Therefore, the GEO/MEO satellites are appropriate to collect the status of LEO satellites and send control commands for routing and resource scheduling. Meanwhile, GEO/MEO satellites are suitable for delivering information without high requirements for the delay, which are already adopted by the O3b system with MEO constellation.

The Earth layer provides services for users while interacting with the GEO/MEO layer. The ground user sends the data to the LEO satellite corresponding to its coverage area and then transmits the data to the target user via the LEO layer. A large-scale LEO constellation covers the Earth, making it easier to communicate in remote areas, oceans, air, and even polar regions. In addition, when the computing resources are limited, the ground computing centers can receive the status of LEO satellites from the GEO/MEO layer to calculate routing and resource allocation and then send them back to the GEO/MEO satellites. Meanwhile, the satellites in GEO/MEO layer are controlled by the ground stations.

 \begin{figure*}
    \centering
    \includegraphics[width=7 in]{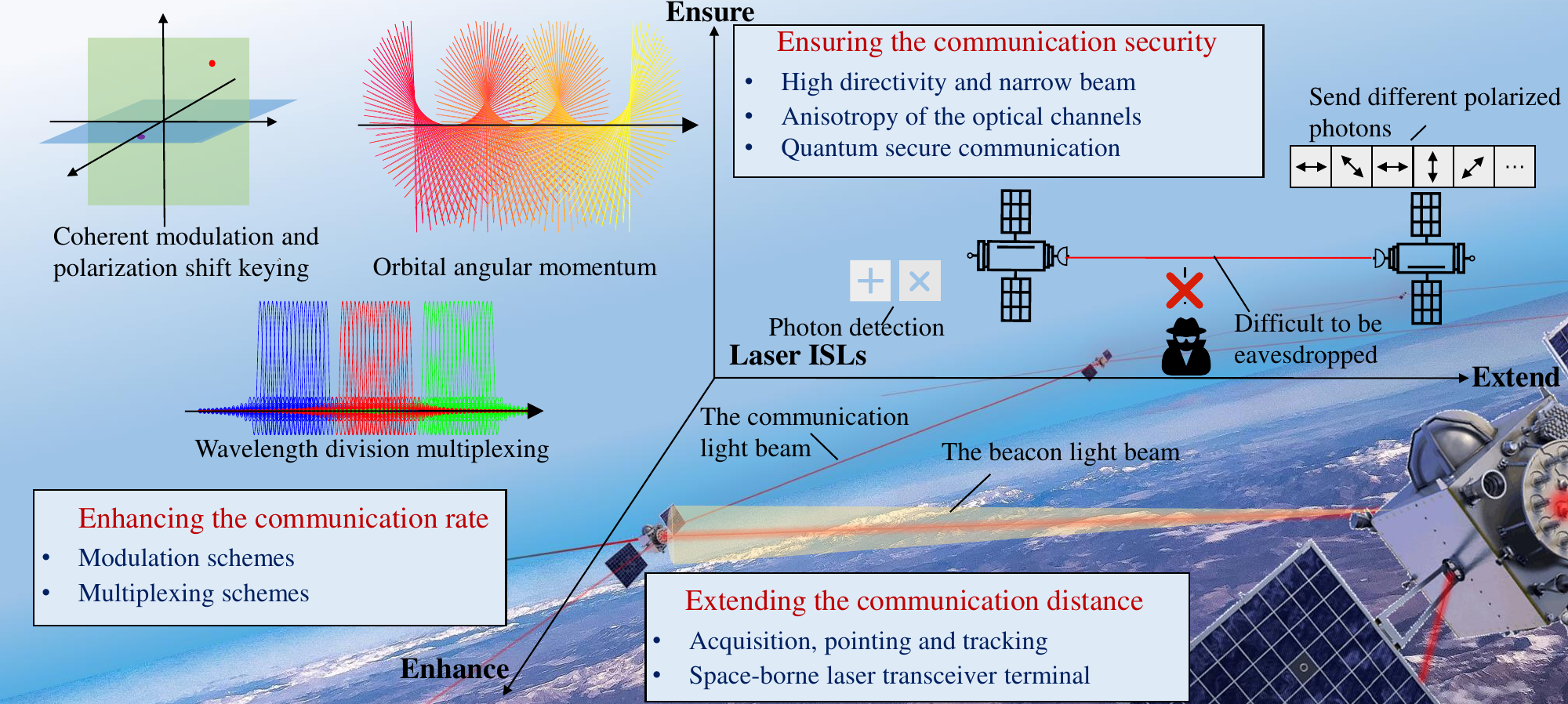}
    \caption{The three advantages of laser ISLs: extending the communication distance, enhancing the communication rate, and ensuring the communication security.}
    \label{advantage}
\end{figure*}

\section{Extending the communication distance}
Based on the aforementioned characteristics of laser signal sources, the merits of laser ISL can be summarized as three points: extending the communication distance, enhancing the communication rate, and ensuring the communication security. As shown in Fig.~\ref{advantage}, the following will elaborate on these advantages respectively.

Because of its high energy concentration and small beam divergence, the laser can meet the link budget of long-distance propagation with lower power than other band, however, this puts higher requirements on the acquisition, pointing and tracking (APT) system of the laser terminals.

\subsection{Acquisition, Pointing and Tracking}
The main task of the APT mechanism is to achieve accurate beam alignment between transmitter and receiver ends and maintain link connectivity when communicating. Due to the large relative motion range between satellites and their mechanical vibration, a coarse pointing assembly (CPA) with a large detection range and a fine pointing assembly (FPA) with high precision are required to maintain the stability of the link. At the same time, a photo-detector (PD) is needed to sense the distribution of received light intensity to determine the incidence angle of the laser, and an inertial measurement unit (IMU) is employed to measure the vibration of the satellite.

In the initial stage of APT, the two satellites determine the general orientation of each other according to the ephemeris and use the beacon light with a wide beam to scan the designated area, in which the CPA is used to align the angle of view. After detecting the beam, the satellites determine the transmitting direction of the receiving beam by the sensors and send a confirmation signal in reverse. After the bidirectional confirmation is completed, the sender switches to a narrow beam for high-rate communication and performs real-time fine-tuning according to the feedback information of sensors and IMU by FPA. Due to the high relative velocity between LEO satellites, a point ahead assembly (PAA) is also utilized to eliminate the effect of relative motion.

For CPA, a large steering range is required, which is achieved by using a gimbaled flat mirror and a telescope, while FPA and PAA require high accuracy and response speed, a fast steering mirror (FSM) is adopted to achieve fast beam steering. Moreover, a quadrant detector or a focal plane array is employed to perceive light intensity distribution and judge the state of the incident beam.
\begin{figure}
    \centering
    \includegraphics[width=3.5 in]{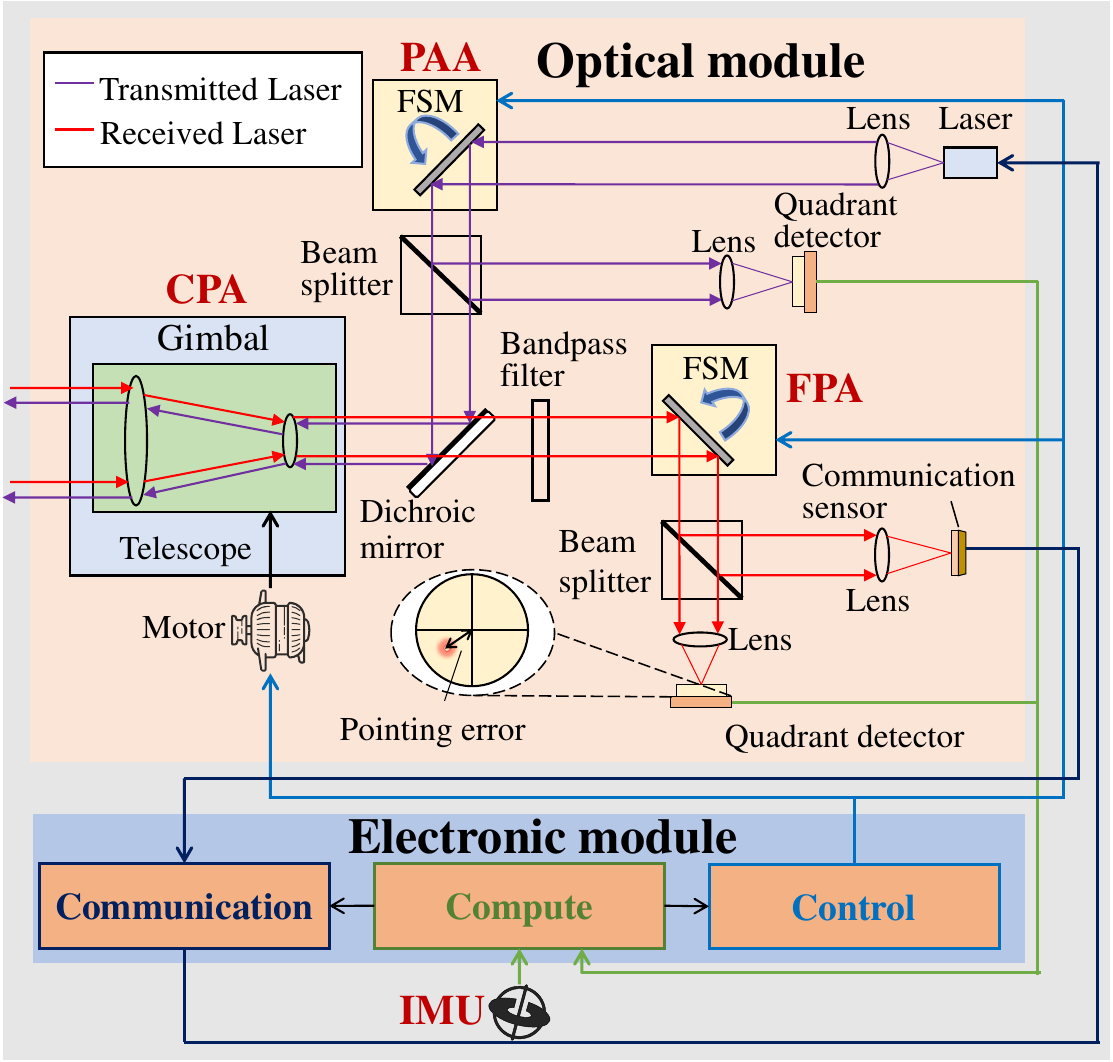}
    \caption{Schematic diagram of space-borne laser terminal.}
    \label{terminal}
\end{figure}
\subsection{Space-borne Laser Transceiver Terminal}

The block diagram of a typical space-borne laser terminal is shown in Fig.~\ref{terminal}. A terminal can be divided into an optical module and an electronic module. The optical module mainly realizes transmitter-receiver isolation and APT mechanism. Different wavelengths are used to receive and transmit signal, and isolated by a dichroic mirror. The received signal is divided into two branches by the beam splitter after the FPA. One is the communication branch, which is converted into electrical signal by the PD and processed by the communication module, while the other enters a quadrant sensor to calculate the pointing error. Likewise, the transmitted signal is divided into two branches after PAA, one for transmitting and the other for measuring deviation. The electronic system is comprised of control, compute, and communication modules. The compute module collects the information of each sensor and IMU for calculation, and transmits the results to other modules. The communication module processes the electrical signal and determines the switch between the beacon light and communication light based on the compute module, and the control module drives the FSM and controls the gimbal by a motor. A terminal called CONDOR designed in ~\cite{laserterminal1} adopts similar architecture, realizing the establishment of laser link within 2 seconds, and communicating at a rate of 5 Gbps with more than 7,000 km distance, in which a steering range of $-175\sim+175$ deg in azimuth and $-25\sim+5$ deg in elevation is achieved. The CLICK terminal~\cite{click} designed for the smaller CubeSats has given up the gimbal architecture and adopts the aircraft attitude pointing to achieve CPA, and so its size is even smaller. The latter design can be used to connect back and forth satellites in the same OP due to the low relative velocity.

\section{Enhancing the communication rate}
Optical ISLs can communicate at a rate of Gbps over thousands of kilometers, which not only benefits by the large bandwidth of the optical band but also profits from the advanced modulation and multiplexing methods.
\subsection{Modulation Schemes}
Due to the high frequency of the optical band, it is difficult to obtain the local oscillator for the coherent detection which has better performance. Adopting the signal intensity to carry information is widely used in optical communication systems. On-off keying (OOK) modulation decides whether to send a pulse within a symbol based on the information bit. The capacity of the direct detection photon channel is investigated in~\cite{ChannelCap}, and it is pointed out that there is a positive correlation between the channel capacity and the symbol peak-to-average power ratio (PAPR). Therefore, pulse position modulation (PPM) with higher PAPR is more efficient than OOK. Meanwhile, PPM also has some derivative types such as differential PPM (DPPM) to further improve its efficiency. Besides, the sub-carrier intensity modulation (SIM) utilizes the phase information by modulating the signal on the intermediate frequency (IF) signal and then using the IF signal to modulate the intensity of the light source. 

\begin{figure}
    \centering
    \includegraphics[width=3.5in]{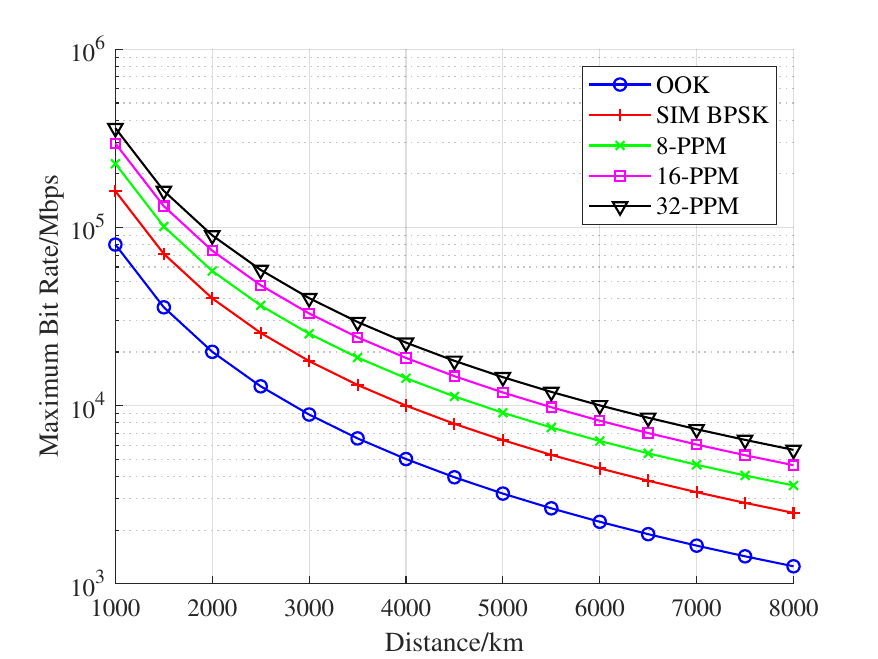}
    \caption{Maximum bit rate under different communication distance with various modulation schemes.}
    \label{bitrate}
\end{figure}

The IM-DD is widely used in space-borne laser communication due to its ease of implementation~\cite{click,laserterminal1}. Fig.~\ref{bitrate} shows the relationship between the maximum transmission rate and the communication distance for the aforementioned OOK, PPM, and SIM BPSK modulation under the link conditions calculated by the data given in~\cite{laserterminal1}, in which the path propagation attenuation, transceiver gain, and energy loss caused by APT error are considered. The scheme with low communication rate has advantages in computational complexity and other aspects, such as OOK has simpler modulation and demodulation and occupies less bandwidth at the same rate than PPM.

In recent years, with the development of optical phase-locked loops and other devices, optical coherent modulation has been fully studied and demonstrated. Moreover, the CCSDS 141.10-O-1 specification suggests a combination of phase and intensity modulation. Optical communication based on coherent modulation has a lower receiving sensitivity requirement, can better suppress background light noise, and has higher spectrum efficiency, which is adopted by the European data relay system~\cite{EDRS}. Moreover, re-using the coherent transceiver from fiber optics to realize the coherent modulation in free space was adopted in TeraByte InfraRed Delivery program~\cite{TBIRD}. Furthermore, taking advantage of the fact that the polarization direction of electromagnetic waves has two degrees of freedom, the information can be carried by the modulation of the polarization direction, which is called polarization shift keying (PolSK).
\subsection{Multiplexing Schemes}
Optical communication can also achieve time division multiplexing (TDM), code division multiplexing (CDM), and frequency division multiplexing (FDM). Moreover, similar to polarization modulation, polarization multiplexing can also be achieved by using two orthogonal polarization directions. Since the TDM, CDM, and polarization multiplexing are straightforward, they will not be repeated here.

Limited by the modulation rate and complexity of the emitters, a single optical signal cannot fully utilize the rich spectrum resources of the optical band. The source with different bands should be used as the central frequency for modulation and they can be distinguished by the wavelength at the receiver. This multiplexing method is called wavelength division multiplexing (WDM), a type of FDM. Although there are unutilized frequency bands between different wavelengths, which sacrifices part of the spectral efficiency, the WDM can take advantage of the large bandwidth of the laser link. Moreover, with the development of optical devices, dense wavelength division multiplexing (DWDM) reduces the interval between different wavelengths to less than 1 nm, which can further improve the utilization rate of the spectrum.

Mode division multiplexing (MDM) can be realized according to the characteristic that the different orbital angular momentum (OAM) modes of photons are orthogonal. The OAM of photons, whose macro representation is a vortex beam, has infinite orthogonal states in theory, which can greatly increase the channel capacity. In recent years, significant progress has been made in the generation and separation of vortex light by Dammann grating. With the combination of WDM and OAM mode multiplexing, the spectral efficiency of more than 20 bit/s/Hz and the transmission capacity of 100 Tbit/s have been achieved~\cite{Tbit}.

\section{Ensuring the communication security}
\subsection{High Directivity and Narrow Beam}
The optical band has a shorter wavelength, and so the beam width can be narrower than other bands with the same aperture of antennas. At the same time, the Gauss beam can be collimated by the beam expander to obtain a smaller divergence. The beam divergence angle of existing space-borne laser terminals is dozens of microradians. In the CONDOR system, which aims to achieve communication over thousands of kilometers, the beam divergence is 17.44 $\rm \mu rad$. It means that more than 80\% of the energy is concentrated in the circular region with a radius of 90 m centered on the aiming center after the laser beam propagates 5,000 km, which makes it difficult to eavesdrop.

Meanwhile, for the receiver, the received intensity decays exponentially with the square of the ratio between the APT deviation and beam width, which means a small alignment error will cause a great attenuation of the received energy. It implies that when both sides complete the APT mechanism and establish a stable communication link, it requires a large amount of energy to interfere with the communication since the interference signal is difficult to coaxial with the transmitter and the receiver. Therefore, the laser beam has an effective resistance to interference.

\subsection{Anisotropy of the Optical Channel}
The optical channel for ISLs is mainly affected by background light and various cosmic rays. Meanwhile, very low Earth orbit satellites operate in the upper ionosphere, and free particles in the ionosphere will also affect the transmission of electromagnetic waves. Because of the directivity of background noise and the overall motion trend of ionospheric particles, there is anisotropy in the optical ISLs of LEO. If the receivers are in different orientations, the channel state information (CSI) will also change. Using the difference of CSI between the legitimate receiver and eavesdropper, the signal quality of the eavesdropper should be reduced without affecting the receiver by reasonable precoding and constellation design, to ensure communication security. 

\subsection{Quantum Secure Communication}
Quantum communication is theoretically secure, which is different from traditional encryption methods that are computationally secure. For photons, the quantum properties are particularly significant in the polarization state. Meanwhile, the laser is generated by stimulated radiation, and its polarization direction is the same as that of the optical pump, and so it is convenient to prepare photons with a specified polarization state. The quantum key distribution (QKD) method has been well studied. The ultimate limits and the practical rates of QKD via satellites under actual free-space channel conditions are discussed in~\cite{QKD}. 

In recent years, the theory of quantum secure direct communication (QSDC) using the quantum entanglement effect has been proposed. The sender transmits only one of each pair of entangled photons at a time, and the measurement for a single photon will make the quantum state of the entangled photons pair collapse, and so it can always confirm whether there is an eavesdropper after a single transmission, then the communication security can be ensured. Since the space channel is simpler than the ground channel, quantum secure communication is easier to realize in satellite communication~\cite{quantumsatellite}.

\section{Future trends and open issues}
\begin{figure*}
    \centering
    \includegraphics[width=7in]{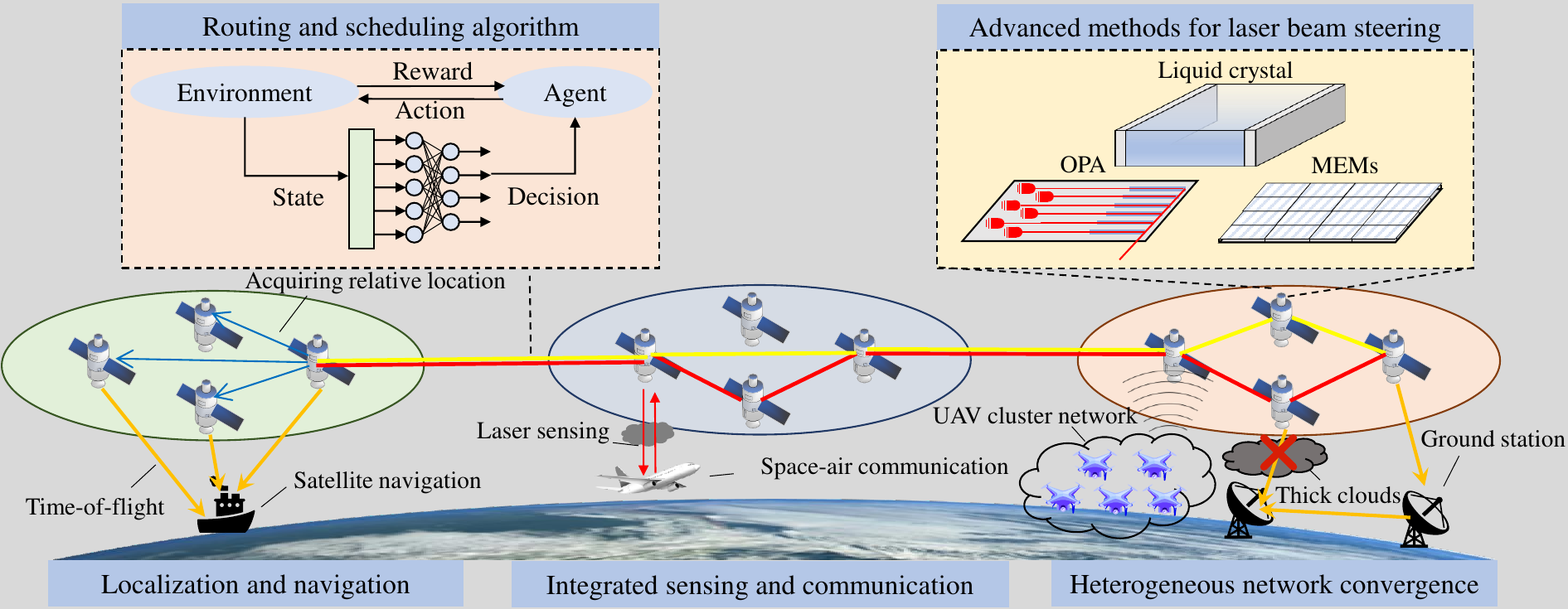}
    \caption{Future trends and open issues for optical ISLs.}
    \label{trend}
\end{figure*}

The optical ISLs have great potential in terms of communication distance, data rate, and communication security, but also have much room for improvement in many aspects as shown in Fig.~\ref{trend}. 

\textbf{Advanced Methods for Laser Beam Steering:}
In the current space-borne terminals, the gimbal for beam steering occupies a large volume and mass. With the development of optical device manufacturing technology, many beam steering schemes are using optical characteristics, including microelectro-mechanical-system (MEMs), optical phased array (OPA), and liquid crystal schemes~\cite{2DMEMS}. The MEMs utilizes a power splitter to divide the beam into several sub-beams and uses micro-motors to drive different optical units, such as mirrors or gratings, to control the direction of the sub-beams, and thus achieving the overall beam steering. OPA uses phase shifters made of electro-optic or thermo-optic materials to form a two-dimensional array. The incident light of OPA also passes through the power splitter and the phase shifting unit, and then through the transmitting unit. The refractive index of the liquid crystal can be changed by applying a periodic stepped voltage to different parts, forming a structure similar to a blazed grating to change the direction of the beam. These technologies can be integrated into silicon chips and are suitable for satellites with limited SWaP, but there are still some deficiencies. The response time of general liquid crystal materials is tens of milliseconds, and the ferroelectric liquid crystal materials with fast response need a higher voltage drive. The beam steering range of OPA or MEMs is also restricted because of the limited scale. Moreover, these schemes are controlled by circuit switches, and so the steering angles are still not accurate enough for tasks with high precision requirements. Larger arrays and hybrid analog-digital drives may be required in the future to improve accuracy.

\textbf{Heterogeneous Network Convergence:}
The satellite communication system cannot meet the communication demands alone, and it should exist as a supplement and enhancement of the existing network. Therefore, it is important to realize the space-air-ground integrated network. Long-distance communication with low delay can be realized by ISLs, and so the heterogeneous network comprised of satellite and terrestrial networks can provide different quality of service (QoS) for various tasks. Moreover, when one part of the heterogeneous network is heavily loaded or unstable, others can provide backups. Algorithms and standards can be designed to implement handover between terrestrial and satellite networks according to QoS and service types. At the same time, the communication terminal should also be compatible with a variety of networks. For the connection between the satellite network and other networks, since laser propagation in the atmosphere is severely degraded by rain and fog, the microwave is generally used for satellite-ground communication. The traditional method needs to translate the optical signal into the electrical signal and then modulate it into the microwave. The emerging microwave photonics can directly convert optical signal to microwave signal, which makes it suitable for the connection with heterogeneous networks such as unmanned aerial vehicle (UAV) cluster networks. Furthermore, a space-air-ground architecture was proposed~\cite{SAGnet} to reduce atmospheric turbulence and increase the usability of optical link.


\textbf{Routing and Scheduling Algorithm:}
Due to the fast movement of low-orbit satellites, the topology structure of the inter-satellite network changes rapidly, which poses a challenge to the communication routing and scheduling algorithm. In recent years, the method to find the optimal policy by machine learning has achieved considerable results in many problems, and it has also been applied to various kinds of resource scheduling algorithms. In the case of the ground-LEO-GEO three-layer constellation model, the throughput of the system is optimized by reinforcement learning, and feature functions are designed based on the characteristics of the model to fit the state-action value function~\cite{RLschedule}. Furthermore, due to the uneven geographical and time distribution of communication demand, it is necessary to sacrifice the delay performance and choose a variety of paths to make the satellite load more balanced. In addition, laser terminals are sensitive to pointing deviation, and need to be powered on all the time to maintain the link, therefore, ISLs with low usage can be turned off to improve energy efficiency. Moreover, the distribution of satellite constellations and ground stations can also be optimized to improve the energy efficiency of communications.


\textbf{Localization and Navigation:}
Because of the large coverage of satellite communication, it is reasonable to combine satellite communication with navigation, measurement, and control services. Moreover, a large number of LEO satellites help reduce the variance of the estimation results. In~\cite{LEOnav}, a navigation method using the Doppler measurements of LEO satellites is proposed, which is different from the traditional way of employing GEO satellites for positioning, making LEO satellites able to communicate and navigate simultaneously. However, for laser transmission, the detection of Doppler shift is complicated, the localization can be achieved by using the time-of-flight of optical signal from different satellites. Furthermore, the localization and navigation have requirements for the precision of location information of the satellites. The approximate location of satellites can be calculated by the orbit parameters, and the precise relative position between satellites can be obtained through the APT mechanism when the ISL is established, and the relative position information between satellites can be used to reduce the error of the satellite position.

\textbf{Integrated Sensing and Communication:}
Based on the similarity between communication and sensing in device architecture and signal processing, integrated sensing and communication (ISAC) has attracted widespread attention. Because the space-borne laser terminals can realize accurate beam steering and acquisition, it is suitable for the ISAC. With a space-borne laser terminal, the outgoing beam is adjusted according to the satellite's motion and the beam stability, and the terrain height is measured according to the return wave~\cite{gaofen}. In addition, satellite laser remote sensing can also achieve the perception of air pollution, cloud thickness, and other information, which is also helpful to sense the link status to improve the quality of communication. Furthermore, the trade-off between communication rate and sensing accuracy can be achieved by flexible waveform design. The pulse-modulated waveform is suitable for sensing, while for communication, the PAPR of signal is constrained to reduce the distortion of the received signal, a reasonable design of the modulated waveform should be considered. Besides, the evaluation indicators of ISAC should be adjusted according to the specific task to meet the various demands for communication and sensing.

\section{Conclusion}
This paper introduces the system architecture of satellite networks and compares ISLs in terms of different methods. The merits of laser ISLs are highlighted from three aspects, the communication distance can be greatly extended by the laser terminal, advanced modulation and multiplexing methods enhance the communication rate, and the communication security benefits from the narrow beam, anisotropy of optical channels and the quantum properties. In addition, the possible development directions of laser ISLs in terms of devices, algorithms, networks and applications are discussed, which can provide some meaningful guidance for future researchers. Laser ISLs will greatly improve the performance of satellite communication and play an important role in the next generation of communication.
\bibliographystyle{IEEEtran}
\bibliography{ref.bib}

~\\
~\\
~\\
~\\
\noindent
\footnotesize{\textbf{Guanhua Wang} is a Ph.D. student with the Department of Electronic Engineering, Tsinghua University, Beijing, China. His research interests are in deep learning for communication and visible light communication.}

~\\
~\\
~\\
~\\
\noindent
\footnotesize{\textbf{Fang Yang} (M’11–SM’13) received the B.S.E. and Ph.D. degrees in electronic engineering from Tsinghua University, Beijing China, in 2005 and 2009, respectively. Currently, he is an Associate Professor with Department of Electronic Engineering, Tsinghua University. He has published over 180 peerreviewed journal and conference papers. He holds over 50 Chinese patents and two PCT patents. His research interests are in the fields of power line communication, visible light communication, and digital television terrestrial broadcasting. He received the IEEE Scott Helt Memorial Award (best paper award in IEEE Transactions on Broadcasting) in 2015. He is a fellow of IET.}

~\\
~\\
~\\
~\\
\noindent
\footnotesize{\textbf{Jian Song} (M’06–SM’10–F’16)  received the B. Eng and Ph.D. degrees in electrical engineering from
Tsinghua University, Beijing, China, in 1990 and 1995, respectively. Currently, he is the Director of Tsinghua DTV Technology R\&D Center. He has been working in quite different areas of fiber-optic, satellite and wireless communications, as well as the power-line communications. His current research interest is in the area of digital TV broadcasting. Dr. Song has published more than 300 peer-reviewed journal and conference papers. He holds two U.S. and more than 80 Chinese patents. He is a Fellow of IET.}

~\\
~\\
~\\
~\\
\noindent
\footnotesize{\textbf{Zhu Han} (S’01-M’04-SM’09-F’14) received his B.S. degree in electronic engineering from Tsinghua University, Beijing, China, in 1997 and his M.S. and Ph.D. degrees in electrical and computer engineering from the University of Maryland, College Park, in 1999 and 2003, respectively. He is a professor in the Electrical and Computer Engineering Department and in the Computer Science Department at the University of Houston, Texas. He is the winner of IEEE Kiyo Tomiyasu Award 2021.}

\balance

\newpage
\vfill

\end{document}